\documentclass[apj]{emulateapj}
\usepackage{graphics,graphicx}
\usepackage{epsfig}
\usepackage{dcolumn}
\usepackage{rotating}
\shorttitle{Q2122-444: a naked AGN fully dressed}
\shortauthors{Gliozzi et al.}
  
  \def\qq{Q2122--444}

  \def\feka{Fe K$\alpha$}
  \def\chandra{{\it Chandra}} 
  \def\xmm{{\it XMM-Newton}} 
   
  \def\atca{{\it ATCA}}

  \def\lum{erg s$^{-1}$}
  \def\flux{erg cm$^{-2}$ s$^{-1}$}
  \def\nh{cm$^{-2}$}
  \def\arcsec{$^{\prime\prime}$}

  \def\ltsima{$\; \buildrel < \over \sim \;$}
  \def\simlt{\lower.5ex\hbox{\ltsima}} 
  \def\gtsima{$\; \buildrel > \over \sim \;$}
  \def\simgt{\lower.5ex\hbox{\gtsima}} 

\begin{document}
\title{Q2122-444: a naked AGN fully dressed}

\author{M. Gliozzi}
\affil{George Mason University, 4400 University Drive, Fairfax, VA 22030}

\author{F. Panessa}
\affil{Istituto di Astrofisica Spaziale e Fisica Cosmica (IASF-INAF), via del Fosso del Cavaliere 100, 00133 Roma, Italy}

\author{F. La Franca}
\affil{Dipartimento di Fisica, Universit\`a degli Studi Roma Tre, via della Vasca Navale 84, 00146 Roma, Italy}

\author{I. Saviane}
\affil{ESO - European Southern Observatory, Alonso de Cordova 3107, Casilla 19001, Santiago 19, Chile}

\author{L. Monaco}
\affil{ESO - European Southern Observatory, Alonso de Cordova 3107, Casilla 19001, Santiago 19, Chile}

\author{L. Foschini}
\affil{INAF - Osservatorio Astronomico di Brera, via E. Bianchi 46, I-23807 Merate, Italy}

\author{L. Kedziora-Chudczer}
\affil{Australia Telescope National Facility, CSIRO, P.O. Box 76, Epping,
NSW 1710 {\it Currently at the School of Physics, UNSW, Sydney, NSW 2052 Australia}}

\author{S. Satyapal}
\affil{George Mason University, 4400 University Drive, Fairfax, VA 22030}

\author{R.M. Sambruna}
\affil{NASA's Goddard Space Flight Center, Code 661, Greenbelt, MD 20771}

\begin{abstract}
Based on previous spectral and temporal optical studies, Q2122-444 
has been classified as a naked AGN or 
true type 2 AGN, that is, an AGN that genuinely lacks a broad line region (BLR). 
Its optical spectrum seemed to possess only narrow forbidden emission lines
that are typical of type 2 (obscured) AGNs, but the long-term
optical light curve, obtained from a monitoring campaign over more than two decades, 
showed strong variability, apparently ruling out the presence of heavy obscuration.  
Here, we present the results from a $\sim$40 ks XMM-Newton observation of Q2122-444
carried out to shed light on the energetics of this enigmatic AGN.
The X-ray analysis was complemented with ATCA radio data to assess the possible
presence of a jet, and with new NTT/EFOSC2 optical spectroscopic data to verify
the actual absence of a BLR. The higher-quality optical data 
revealed the presence 
of strong and broad Balmer lines that are at odds with the previous 
spectral classification of this AGN. The lack of detection of radio emission rules out
the presence of a jet. The X-ray data combined with simultaneous UV
observations carried out by the OM aboard \xmm\ confirm that Q2122-444 is a typical
type-1 AGN without any significant intrinsic absorption. New estimates of the black
hole mass independently obtained from the broad Balmer lines and from a new scaling
technique based on X-ray spectral data suggest that \qq\ is accreting at a
relatively high rate in Eddington units.
\end{abstract}


\section{Introduction}
More than two decades of multi-wavelength observations and theoretical studies
have led to a generally accepted unification scheme for AGNs. According to this
scheme, all AGNs share some basic ingredients: 1) a supermassive black hole; 
2) an accretion disk coupled with a hot corona, which radiates from the 
optical through X-ray energies; 3) high velocity and high density 
gas located at $\sim$ pc scales, usually referred to as broad-line
region (BLR), since its radiation is dominated by broad 
(FWHM$\sim$ 2000--10,000 km ${\rm s^{-1}}$) permitted lines;
4) lower velocity and low density gas 
located at $\sim$ kpc scales, referred to as narrow-line region (NLR),
and emitting narrow (FWHM$<$ 1000  km ${\rm s^{-1}}$) 
forbidden and permitted lines. 
A gaseous and molecular torus on pc scales presumably obscures
the inner engine and the BLR when viewed edge-on. In addition, in 
$\sim$ 10\% of AGNs, there are relativistic jets 
extending from kpc up to Mpc scales. Within
this scheme, several observational differences in the broad-band
and line spectral properties between type 1 (objects with broad permitted 
lines) and type 2 AGNs (with narrow lines only; see \citealt{oster81}) are 
explained by orientation effects \citep[e.g.][]{anto93,urrypad95}.

Although observations generally support orientation-based unification models
for AGNs, exceptions do exist \citep[e.g.][]{pappa01,panes02,bian08,trum09}. 
Observationally, only about 50\% of the brightest 
Seyfert 2s show the presence of hidden broad line regions 
\citep[HBLRs;][]{tran01}
in their polarized optical spectra. Several studies 
suggest that the presence or absence of HBLRs depends on the AGN luminosity, 
with HBLR sources having larger optical luminosities \citep[e.g.][]{tran01,gu02}.
From the theoretical view point, \citet{nica00} hypothesized that the 
absence of HBLRs corresponds to low values for accretion rate onto the central 
black hole. Similar conclusions are reached by \citet{eli09} in the
context of the disk-wind scenario for low accretion rate AGNs.
Further support for the existence of AGNs lacking a BLR was put forward by 
\citet{hawk04}, who reported the discovery of six naked AGNs 
-- a new class of AGNs 
characterized by the absence of the broad line region accompanied by strong 
continuum emission and strong variability in the optical band. This result is
based on a large-scale monitoring program carried out in different optical 
bands over the last 25 years, and on the spectroscopic follow-up (see \citealt{hawk04} 
and references therein for a more detailed description of the optical 
observations).

It must be outlined that, although the
unification models are able to explain most of the AGN observational
properties, they do not offer any insight into the physical origin of the
basic AGN ingredients, namely the BLR, the NLR, accretion flow, and the torus.
Detailed investigations of AGNs without BLR have the potential of shedding
some light on the physical mechanisms that lead to the suppression 
(and possibly also to the formation) of the BLR.
In addition, in the framework of clumpy torus models \citep[e.g.][]{nenko02},
there is growing evidence that the torus and the BLR are not
separate entities. Therefore, the  study of naked AGNs may provide useful
constraints also on the formation mechanism and nature of the torus. Finally,
if indeed naked AGNs lack BLR and strong intrinsic absorption,  they
offer the best opportunity to investigate the accretion flow.
An additional (and perhaps the most important) reason  
is that AGNs without BLR might be representative of a large 
class of AGNs and thus a detailed study of the naked AGNs is critical.
Indeed, recent studies of \citet{stef03}, based on 
Chandra deep field data, indicate that a 
sizable fraction of the local AGN population is without BLs and 
in fact represents the dominant AGN class for X-ray luminosities below 
$\sim 4\times 10^{43}{~\rm erg~s^{-1}}$ \citep[see also][]{law82,lafra05}.
This is consistent with the findings of several recent studies
\citep[e.g.][]{bia07,diam09,plot10} 
that suggest that AGNs without 
BLR or with weak optical lines are associated not only with low-luminosity AGNs 
but also with relatively bright AGNs.

One of the most effective ways to investigate the properties of AGNs is with 
X-ray observations, since X-rays are produced and reprocessed in the 
inner, hottest nuclear regions of the source, and their penetrating power  
allows them to carry information from the central engine 
without being substantially affected by absorption, and to probe the presence 
of the putative torus.

Recently, we carried out a preliminary X-ray study of three naked AGNs 
(Q2122-444, Q2130-431, Q2131-427) with \chandra\ snapshot observations 
that were performed in December 2005. The most relevant results from 
the \chandra\ study are reported in \citet{glioz07} and can be summarized as
follows:
(1) The three AGNs are easily detected in X-rays and appear to be point-like
at the sub-arcsecond resolution of \chandra. (2) Their X-ray spectral 
properties are consistent with those of Seyfert 1 galaxies, with 
typical photon indices of $\sim$1.8 and without strong  intrinsic absorption
($N_{\rm H}<5\times 10^{21}$ \nh).
(3) The three sources are fairly bright in the X-rays with values
comparable with those observed in Seyfert 1 galaxies 
and larger than typical Seyfert 2 
luminosities \citep{turn97}, and  have Eddington
ratios ranging between $10^{-4}$ and  $10^{-2}$ .

More recently, Q2131-427 and Q2130-431, two of the original naked AGNs
detected by \citet{hawk04},  have been 
observed quasi-simultaneously in the X-rays with \xmm\ and 
in the optical band with the NTT by \citet{panes09}. The
high-quality optical spectroscopic data confirmed the ``naked" nature of 
Q2131-427 but revealed weak broad lines for Q2130-431. On one hand, these 
results combined with the lack of intrinsic absorption inferred from \xmm\ led 
\citet{panes09} to the conclusion that Q2131-427 is a ``true" type 2 AGN
and Q2130-431 might be a ``true" intermediate Seyfert with an intrinsically 
weak BLR.  On the other hand, the detection of broad lines in the optical 
spectrum of Q2130-431  
casts some doubts on the optical classification of the original sample of 
\citet{hawk04}, since it is based on relatively low-sensitivity 
optical spectral data. 

Here we investigate the nature of the central engine in \qq\ 
(RA$_{\rm J2000}=21^h26^m03.94^s$, Dec$_{\rm J2000}=-44^o  11' 19"$, $z$=0.311),
another putative naked AGN from the sample of \citet{hawk04}.
To this end we make use of \xmm\ EPIC and OM data (described in Section 3)
to shed light on the energetics of this enigmatic AGN. The \xmm\ analysis is
complemented with the investigation of new optical spectroscopic data 
recently obtained from the NTT/EFOSC2 to robustly verify
the claim that \qq\ lacks a BLR (see Sect. 4). We also use radio data from the 
Australia Telescope Compact Array (ATCA)  to assess whether the peculiar properties
of this source can be explained by 
the presence of a jet. The description of all the observations 
and the data reduction is provided in Section 2. The main results and their
implications are discussed in Section 5, while the main conclusions
are summarized in Section 6.

\section{Observations and Data Reduction}

\subsection{XMM-Newton}
\qq\ was observed by \xmm\ from November 18, 2007, 22:38 UT to November 19,
2007, 10:22 UT. The EPIC pn and MOS cameras were operated in full-frame mode 
with a thin filter. The data were processed using the latest CCD gain
values. For the temporal and spectral analysis, events corresponding to pattern 
0--12 (singles, doubles, triples, and quadruples) in the MOS cameras and 
0--4 (singles and doubles only, since the pn pixels are larger) in the pn 
camera were accepted. Arf and rmf files were created with the \xmm\ Science 
Analysis Software (\verb+SAS+) 8.0.
The recorded events were screened to remove known hot pixels and data affected
by periods of flaring background. The resulting effective total exposures are
34 ks for the EPIC pn and 41 ks for both MOS cameras. The extraction radius used 
for the source spectra and light curves is 30\arcsec.
Background spectra and light curves were extracted from
source-free circular regions of 60\arcsec\ extraction radius
on the same chip as the source.  There are no
signs of pile-up in the pn or MOS cameras according to the {\tt SAS}
task {\tt epatplot}.  
The RGS data  have signal-to-noise ratio ($S/N$) that is too low for a 
meaningful analysis.

The X-ray spectral analysis  was performed using the {\tt XSPEC v.12.4.0}
software package \citep{arn96}.  The EPIC data have been re-binned in order 
to contain at least 20 counts per channel for the $\chi^2$ statistic to be valid. 
The errors on spectral parameters are at 90\% confidence level for one 
interesting parameter ($\Delta \chi^2=2.71$).

The data from the OM \citep{mas01} were processed with the latest 
calibration files  using the {\tt SAS}
task {\tt omichain}, which provides  count rates in the B, U,
UVW1, and UVM2 bands. The count rates were then
converted into magnitudes using the formula 
${\rm mag=-2.5\log(count~rate)+zeropoint}$ (where the zero-points are 19.266, 18.259,
17.204, and 15.772, for the B, U, UVW1, and UVM2 bands, respectively)
and corrected for systematics. The values were 
corrected for extinction (see Sect 3.3 for
more details) and converted to fluxes 
by using standard formulae \citep[e.g.][]{zom90}.

\subsection{NTT/EFOSC2}
The optical spectroscopic data of \qq\ were retrieved from the ESO archive.
\qq\ was observed with ESO Faint Object Spectrograph and Camera v.2 (EFOSC2) 
at the New Technology Telescope (NTT) in La Silla (Chile) on 2009, July 6th
with grating \#13 and a 1-arcsec slit width which corresponds to a wavelength resolution of 21.2 \AA\ (FWHM), in the
range 3900-9200 \AA.
Three spectra with exposure time of 600s each were acquired (air mass $\sim$ 1.08).  
A standard reduction process was applied using midas and iraf tasks. The raw data
were bias subtracted, cleaned from cosmic rays and corrected for pixel-to-pixel variations (flat-field). Object spectra were extracted
and sky subtracted. Eventually the three spectra were averaged to obtain a single higher S/N spectrum.
Wavelength calibrations were carried out by comparison with exposures of He and Ar lamps, with an accuracy 
$<$ 15 \AA. Relative flux calibration was carried out by observations of several spectrophotometric standard stars (Oke 1990).

\subsection{ATCA}
We used the ATCA in the 6A configuration for imaging \qq\  at 4.8 and 8.6 GHz.
The beam size at 8.6 GHz is 1.3$\times$0.8\arcsec\ with a position angle of 78.9$^o$. 
At 4.8 GHz the beam size is 2.2$\times$1.4\arcsec\ with a position angle of 58.9$^o$. Typically
the positional accuracy may be affected by pointing stability of the array. However, the
pointing deviations are usually below 1 arcsec even for the snapshot observations.
The source was observed in three sessions over the period between 20 and 22 Jan 2008. 
In each session we obtained 
six 10-minutes scans with 1 second integration time at each frequency, which were interleaved with the observations of the point source phase calibrator, PKS 2106-413.
The flux density calibration was achieved with the ATCA primary calibrator, PKS 1934-638. The final images of the 30 (15) arcsec area for 4.8 (8.6) GHz around the optical 
position of the source show no detection of radio emission above the 0.78 mJy noise level limit
(which is the 3 sigma flux limit).

\section{The XMM-Newton view of \qq}

\subsection{X-ray temporal analysis of \qq}
\begin{figure}
\begin{center}
\includegraphics[bb=12 60 440 405,clip=,angle=0,width=9cm]{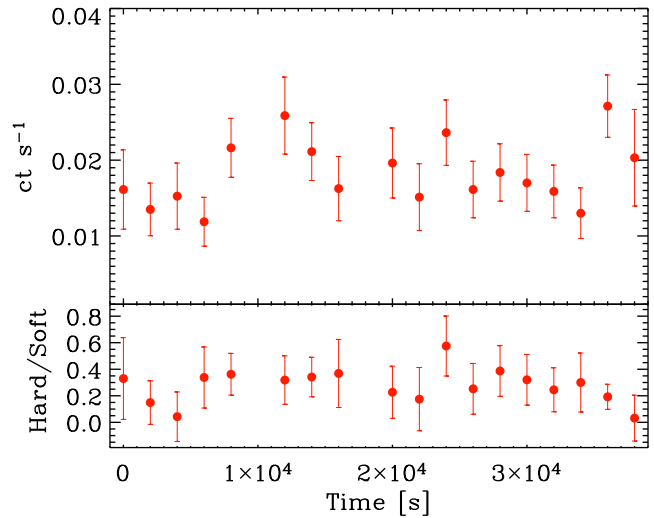}
\end{center}
\caption{\xmm\ EPIC light curve in the 0.3--10 keV (top panel),
and HR=(2--10 keV/0.3--2 keV) plotted versus the time (bottom panel) for \qq. }
\label{figure:fig1}
\end{figure}

We studied the short-term variability of \qq\ using EPIC pn data with time-bins
of 2000 s. Figure~\ref{figure:fig1} shows the EPIC time series of the 0.3-10 keV
band (top panel) and of the hardness ratio $HR=$2-10 keV/0.3-2 keV (bottom panel).
A visual inspection of Fig.~\ref{figure:fig1} suggests that low-amplitude flux changes 
might be present in the light curve without being accompanied by spectral variations.
A formal analysis based on a  $\chi^2$ test indicates that there is no significant
temporal ($P_{\chi^2}\simeq 32\%$) or spectral ($P_{\chi^2}\simeq 92\%$) variability
on short timescales.

Since \qq\ has also been observed by \chandra\ in December 2005, it is possible to
investigate the presence of long-term flux variations. Over a period of nearly
two years the 2--10 keV flux of \qq\ decreased by a factor of $\sim$2 (from
$7.4\times10^{-14}$ \flux\ measured during the \chandra\ snapshot survey
to $3.9\times10^{-14}$ \flux\ measured by \xmm). Interestingly, this
significant change of X-ray flux appears to be accompanied by a moderate spectral 
variability: the hardness ratio of \qq\ increases from $HR_{\rm Chandra}=0.27\pm0.11$ to 
$HR_{\rm XMM}=0.60\pm0.04$. This suggests that \qq\ follows the typical spectral 
variability trend observed in Seyfert 1 galaxies, with a steeper X-ray spectrum when the source is brighter \citep[e.g.][]{sobo09}.

\subsection{X-ray spectral analysis of \qq}
\begin{figure}
\includegraphics[bb=40 2 567 704,clip=,angle=-90,width=9.cm]{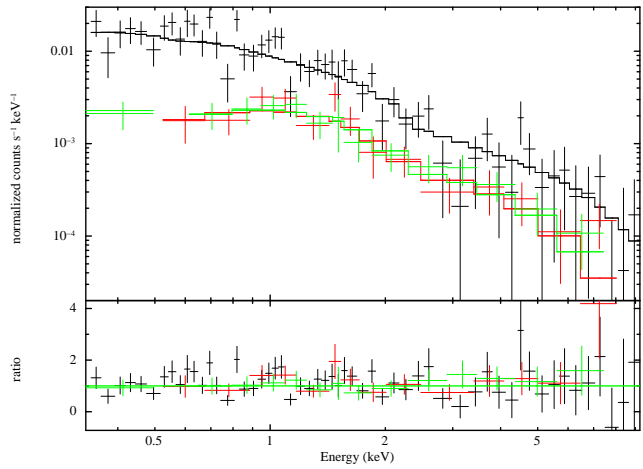}
\caption{EPIC spectra of Q2122-444 and data/model ratios to a simple power-law model modified by photoelectric absorption. Black, red, and green symbols represent the EPIC pn, MOS1, and MOS2 data,
respectively.}
\label{figure:fig2}
\end{figure}

\begin{figure}
\includegraphics[bb=110 90 540 655,clip=,angle=-90,width=9.cm]{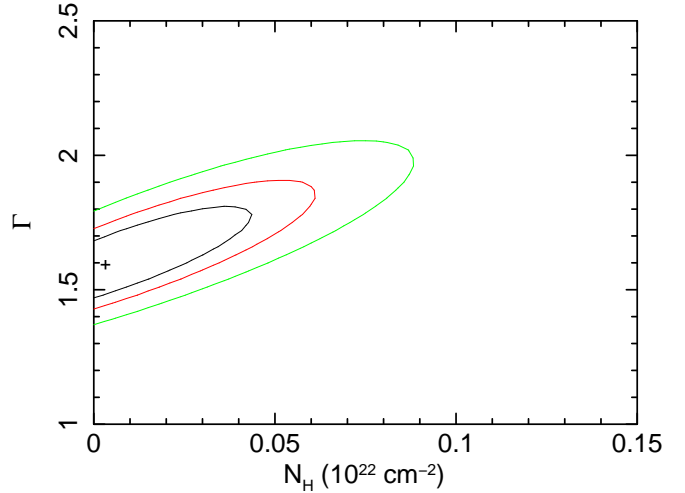}
\caption{Confidence contours (68\%, 90\%, and 99\%) in the $\Gamma$ --
$N_{\rm H}$ plane for Q2122-444.}
\label{figure:fig3}
\end{figure}

Unlike the \chandra\ spectral analysis that was severely hampered by poor 
photon statistics due
to the brevity of the observation, the longer \xmm\ exposure coupled with
its high throughput makes it possible to study the X-ray spectral properties of
\qq\ in greater detail. However, similar to the \chandra\ ACIS spectrum,
the combined spectra of the three EPIC cameras are adequately fitted 
by a simple power law modified by absorption that  is left free to vary
($\chi^2_{\rm red}=1.05$; 112 d.o.f). 
The best-fit values of the relevant spectral parameters are: 
$\Gamma=1.62\pm0.14$ and $N_{\rm H}=8_{-8}^{+23}\times10^{19}$ \nh.
Similar values of $\Gamma$ and $\chi^2_{\rm red}$ are obtained when
the column density is fixed at the Galactic value $3.6\times10^{20}$ \nh.
The spectra and residuals of Q2122-444  are shown in Figure~\ref{figure:fig2}, while
Figure~\ref{figure:fig3} shows the confidence contours of $\Gamma$ and  $N_{\rm H}$.
 
It is instructive to compare these results with the corresponding ones
obtained with the \chandra\ snapshots (see Figs. 2 and 3 from 
\citealt{glioz07}). Although the parameters from the \xmm\ analysis are 
more tightly constrained, the basic conclusions remain the same: the photon 
index is consistent with typical values measured in Seyfert 1 galaxies and,
more importantly, there is no evidence for intrinsic absorption (the 90\%
confidence contours are fully consistent with the Galactic value, 
$N_{\rm H,Gal}=3.6\times10^{20}$ \nh).

Although the EPIC cameras have a larger effective area above 6 keV with respect to
the \chandra\ ACIS camera, the limited statistics of the \xmm\ spectrum of \qq\ does not
allow a detailed investigation of the \feka\ line. The line energy and width 
remain unconstrained when they are left free to vary, therefore we fixed them at
$E=6.4$ keV and $\sigma$=0.01 keV, respectively.
The model comprising a \feka\ line in addition to the absorbed power law 
does not lead to any statistically significant improvement of the fit 
($\Delta\chi^2\simeq 1$ for one additional parameter) indicating that a \feka\
line is not required. The equivalent width of this line is 640 eV with a 90\%
upper limit of $\sim$1 keV.

Since the X-ray emission is likely to be produced by the Comptonization 
of seed photons emitted by the underlying accretion disk, it is appropriate 
and physically more motivated to use a Comptonization model instead of a
phenomenological model like the power law. Using a general Comptonization model,
the Bulk Motion Comptonization  
(BMC; \citealt{tita97}), yields a fit ($\chi^2_{\rm red}=1.08$; 110 d.o.f) which 
is statistically indistinguishable from the power law fit, with parameters
that are fairly typical for AGNs. Specifically, we obtain
$kT=0.1$ keV (the temperature of the seed photons), $\alpha=0.63\pm0.14$ (the spectral
index, which is related to the photon index by $\Gamma=\alpha+1$),
$\log A=0.11$ (which is related to the Comptonization fraction $f$  
by the relation $f=A/(1+A)$, where $f$ is the ratio of Compton 
scattered photons over the seed photons), and the normalization
$N_{\rm BMC}=(2.8\pm0.2)\times 10^{-7}$, which is
in units of $(L/10^{39}\; {\rm erg~s^{-1}})(10\;{\rm kpc}/d)^2$ with
$L$ and $d$ being the luminosity and distance of the object,
respectively. 

The use of the BMC model is the 
first step necessary to constrain the black hole mass in \qq. The estimate 
$M_{\rm BH}$ is obtained from the X-ray spectral results of the BMC model
by making use of a scaling technique recently introduced for Galactic black holes 
by \citet{shap09} and extended to AGNs by \citet{glioz10}. This procedure
is described in Section 5.

\subsection{The OM view of \qq}
The simultaneous use of several cameras and telescopes on board of \xmm\ is one 
of the most beneficial capability of this satellite. This property not only makes
it possible to increase the S/N in the X-ray band by combining the data of the 
three EPIC cameras, but thanks to the OM it also offers the simultaneous coverage 
of the optical and UV bands, which is crucial to understand the energetics of AGNs.

During the \xmm\ pointing in November 2007, the OM detected \qq\ in the 
B (4400 \AA), U (3440 \AA), and UVW1 (2910 \AA) bands, whereas the source
was not detected with the UVW2 (2120 \AA) filter. The observed magnitudes
are respectively: $m_B=21.1\pm0.4$, $m_U=20.2\pm0.2$, $m_{UVW1}=20.1\pm0.2$.
Since there is no general consensus on the extinction 
corrections that need to be applied in the optical-UV band,
we have tried different extinction curves, including
the Galactic extinction proposed by  \citet{card89}  
and the Small Magellanic
Cloud (SMC) type extinction \citep{prev84}, which proved to
work well on a large sample of AGNs  from the Sloan Digital Sky Survey 
\citep{hopk04}. For completeness, we also applied extinction 
corrections derived from the X-ray spectral fits by converting
$N_{\rm H}$ into optical extinction with the relation $A(V)/N_{\rm H}=
5.3\times 10^{-22}$ \citep{cox00}. The results from 
the different extinction laws are consistent with each other 
within 10\%.

One of the most widely used quantities to parametrize the broadband properties
of AGNs is the the spectral index, $\alpha_{\rm OX}= -
\log(l_{\rm 2500\AA}/l_{\rm 2keV})/\log(\nu_{\rm 2500\AA}/\nu_{\rm 2keV})$
\citep{tanan79}.
We derive $\alpha_{\rm OX}=1.31\pm0.06$, from the simultaneous X-ray and UV
(UVW1 filter) fluxes, assuming a typical UV slope of 0.7 ($f_\nu\propto\nu^{-0.7})$
and the extinction prescriptions discussed above. 
In Section 5 we discuss the implications of this result in the context of 
type 1 AGNs.

\section{Optical Spectroscopy of \qq}

\begin{figure}
\begin{center}
\includegraphics[bb=30 3 595 750,clip=,angle=-90,width=9.cm]{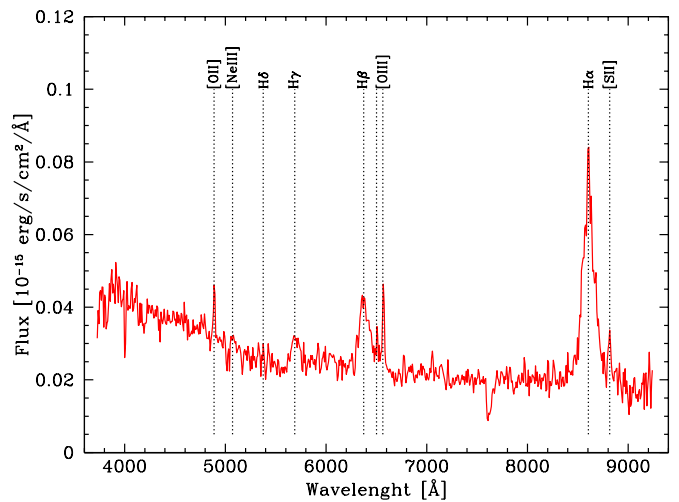}
\end{center}
\caption{ESO NTT/EFOSC2 optical spectrum in the observer's frame. The main emission line 
wavelengths are indicated.}
\label{figure:opt}
\end{figure}

In Figure~\ref{figure:opt} we show the EFOSC2 optical spectrum of \qq.
The line intensities and widths were measured by fitting Gaussian profiles and the Full Width Half Maximum (FWHM)
calculated taking into account the spectrograph resolution. The results from the spectral fitting
of the relevant lines are shown in Table~\ref{table=opt_spec}.

The presence of strong broad Balmer lines clearly points to a reclassification of \qq\
 as a type 1 Seyfert galaxy.
Hawkins (2004) measured a FWHM of 351 km/s and 699 km/s for the H$_\beta$ and $\mathrm{[{O\,\textsc{iii}}]}$ emission lines
respectively (the H$_{\alpha}$ line was out of the covered spectral range) and classified this source as a type 2 AGN.  A possible explanation for the mis-match
between the two optical classifications is that the H$_{\beta}$ broad component may have been obscured by a nearby absorber at the time of the  Hawkins observation,
in agreement with the findings of \citet{risal09}  that demonstrate that variations of the local absorber can occur on short timescales and change the appearance of the central
engine. In addition, it must be pointed out that Hawkins' optical spectroscopy
was performed with a 2dF multi-fibre spectrograph at the AAT which cannot guarantee 
high S/N ratio spectra because of difficulties in subtracting the sky background
\citep[e.g.][]{lissa94,wats98}, and this problem is more severe
for relatively faint objects like \qq.

\begin{table}
\caption{\bf Measured line parameters for the optical spectrum of \qq.}
\small{
\begin{center}
\begin{tabular}{lrrr}
\hline
\hline
\\
\multicolumn{1}{c}{Line} &
\multicolumn{1}{c}{Flux} &
\multicolumn{1}{c}{FWHM} &
\multicolumn{1}{c}{EW} \\
\\
\multicolumn{1}{c}{} &
\multicolumn{1}{c}{($10^{-15}$ cgs)} &
\multicolumn{1}{c}{(km s$^{-1}$)} &
\multicolumn{1}{c}{(\AA)} \\
\\
\hline
\hline
\\
H$_\alpha$    		                        			& 7.87$\pm$0.12     & 4236$\pm$116 & 324.8$\pm$6.5   	\\   
H$_\beta$	 			    			& 2.88$\pm$0.13     & 5166$\pm$129 & 100.4$\pm$4.9     \\
H$_\gamma$ 		 			    		& 1.13$\pm$0.05 & 5660$\pm$760  & 39.4$\pm$1.8  	\\
$\mathrm{[{O\,\textsc{iii}}]}$ 				& 0.59$\pm$0.04 & 898$\pm$22      &  19.4$\pm$1.9  	 \\
$\mathrm{[{O\,\textsc{ii}}]}$ 				& 0.53$\pm$0.43      & 1042$\pm$88  & 14.2$\pm$1.4	 \\
\\
\hline
\end{tabular}
\end{center}
Notes:  Line flux in 10 $^{-15}$ ergs cm$^{-2}$ s$^{-1}$, FWHM in km s$^{-1}$ and rest-frame Equivalent Width (EW) in \AA\  for the 
broad components of the Balmer lines, [OIII] and [OII].} 
\label{table=opt_spec}
\end{table}

\section{Discussion}
The primary goal of our original \xmm\ project was to shed light on the nature
of \qq, a member of the enigmatic class of naked AGNs that allegedly lack a 
BLR based on their unusual combination of spectral and temporal properties 
in the optical band \citep{hawk04}.
The high-quality spectroscopic data presented in the previous 
section revealed the presence of strong broad
lines in the optical spectrum of \qq, ruling out its naked nature.
Nevertheless, it is interesting to investigate the energetics of \qq. 

\subsection{Jet Contribution}
Before discussing the energetics of \qq, it is important to establish
whether a relativistic jet is present in this system and if its energetic 
contribution is important beyond the radio band. The original reason to look
for a jet in \qq\ was that it represents the simplest solution to explain the 
unusual combination of spectral and temporal properties shown by \qq\ in the
optical band. In this scenario, the jet
not only may produce the observed large-amplitude variability of the continuum
but it can also contribute to the non-detection of broad optical lines through
dilution. Additionally, the BLR emission could be partially obscured by a 
local absorber. However, the NTT/EFOSC2 optical spectroscopy has revealed that a BLR 
is indeed
present in \qq\ and the X-ray spectral analysis has confirmed the lack of
relevant intrinsic absorption. Nevertheless it is instructive to investigate 
the radio properties of this source to test the contribution of a putative jet. 

The lack of radio detection in the ATCA observations  
leads to a 3 $\sigma$ upper limit of 0.78 mJy for the radio flux of \qq. 
This value combined with the flux measurements in the optical and
X-ray bands yields low upper limits for the different radio-loudness 
parameters defined in the literature. For example, using the standard
definition of radio loudness \citep[e.g.][]{kell94}, we get 
$R\equiv L_\nu(5~{\rm GHz})/L_\nu({\rm B}) < 5.5$, which is well below
the threshold $R=10$ that is traditionally used to divide radio-quiet
from radio-loud AGNs. Note however that
$R=10$ does not represent a reliable boundary between radio loud (RL) and
radio quiet (RQ) AGNs  \citep{hopeng01}.
A more reliable conclusion can be derived by using 
$R_{\rm x}\equiv\nu L_\nu(5~{\rm GHz})/L_{\rm 2-10~ keV}$, which was 
introduced by \citet{tera03} to avoid extinction problems, 
and exploiting the results of \citet{panes07}, 
who carried out a detailed investigation of  the 
radio loudness in two large samples of radio-quiet 
Seyfert galaxies and Low-Luminosity Radio Galaxies (LLRGs)
and derived a RL/RQ threshold  of $\log R_{\rm x}=-2.755\pm 0.015$.
The upper limit obtained for \qq, $R_{\rm x}<10^{-4}$, indicates
that this source is fully consistent with radio-quiet Seyfert 
galaxies and
inconsistent with radio-loud objects (see Figure 4 of \citealt{panes07}). 
Since we have inferred that either a jet is absent in \qq\ or its 
contribution is completely negligible, for the remainder of the paper 
we can safely assume that the broadband emission is associated with the 
accretion flow only. 

\subsection{Black Hole Mass}
The first necessary step to investigate the central engine in AGNs is 
to constrain the mass of the supermassive black hole.
As with many AGNs for which there are no direct measurements from reverberation
mapping, a BH mass of $\sim 2.2\times 10^8 ~M_\odot$ \citep{glioz07} was determined
using the virial theorem and  FWHM([O III]), which however was proven to be not a 
very robust indicator of the stellar velocity dispersion \citep{netz07}. The broad
Balmer lines detected in the EFOSC2 spectrum in combination with the simultaneously
measured flux at 
5100 \AA, $f_{5100 \AA}=(0.206\pm0.021)\times 10^{-16}$ \flux\ \AA$^{-1}$, allows 
a more reliable estimate of $M_{\rm BH}$. Plugging the value of FWHM(H$_{\alpha}$)  
in the virial formula and using the relation between the broad line region radius 
and the 5100 \AA\ luminosity from \citet{bentz} to determine $R_{\rm BLR}$, we
derive $M_{\rm BH}=(7.1\pm1.4)\times 10^7 ~M_\odot$, where a 20\% error was
conservatively adopted to account for the uncertainties associated with this method.
Slightly larger values are obtained when 
using FWHM(H$_{\beta}$) and FWHM(H$_{\gamma}$) instead of FWHM(H$_{\alpha}$): 
$M_{\rm BH}\simeq (9-10)\times 10^7 ~M_\odot$.

An alternative way to constrain the $M_{\rm BH}$ of \qq\ is based on the analogy
between AGNs and Galactic black holes (GBHs). There is now mounting evidence that AGNs 
may be considered as large-scale analogs of GBHs 
\citep[e.g.][]{merlo03,falk04,koerd06,mchar06}.
Despite the large difference in scales, 
both GBHs and AGNs are believed to harbor the same central engine: a black hole and an 
accretion disk/corona that sometimes produces a relativistic jet. It is also generally 
accepted that the bulk of the X-ray emission in any black hole system, irrespective of
their mass, is produced by the Comptonization process, although some controversy still exists
about the origin of X-rays during the low-hard state in GBHs \citep{zdz03,mrk05}.
Recently, \citet{shap09}
discovered that GBHs during their spectral transition between low-hard to high-soft state
present a universal scalable relationship between the 
photon index $\Gamma$ and the normalization of the BMC model $N_{\rm BMC}$,
and that the $\Gamma - N_{\rm BMC}$ relationship  can be used to estimate the mass of 
any GBHs. 

The physical basis of these scaling techniques is thoroughly explained by \citet{shap09}
and essentially relies on the fact
that the BMC normalization is a function of 
luminosity and distance ($N_{\rm BMC}\propto L/d^2$) and that the luminosity 
of a black hole system can be expressed as $L\propto \eta M_{\rm BH} \dot m$, 
where $\eta$ is the radiative efficiency and $\dot m$ the accretion rate in 
Eddington units. The self-similarity of the $N_{\rm BMC}$ - $\Gamma$ correlation
shown by GBHs implies that, in the same spectral state, different sources have 
similar values of $\eta$ and $\dot m$, and the photon index is a reliable 
indicator for the source's spectral state. 

Since the Comptonization process producing the X-ray emission appears
to be ubiquitous in black holes systems, and since there is evidence
that the photon index is positively correlated with the accretion rate
also in AGNs \citep[e.g][]{shem06,papa09},
in principle, the same method may be applied to estimate the mass of 
supermassive black holes. 

We have recently used this method to constrain $M_{\rm BH}$ 
in the particular case of the Narrow-line Seyfert 1 galaxy PKS 0558-504 
(see \citealt{glioz10}
for details) and we are in the process of testing the extension of this method 
to SMBHs in general by applying it to a sizable sample of AGNs with masses 
determined via reverberation mapping (Gliozzi et al. in preparation).
The basics steps of this scaling method can be summarized as follows:\\
\noindent 1) Construct a $\Gamma -
N_{\rm BMC}$ plot for a GBH of known mass and distance, which will be
used as reference (hereafter denoted by the subscript {\it r}).\\
\noindent 2) Compute the normalization ratio between the target of interest
(that is denoted by the subscript {\it t}) and the reference object
$N_{\rm BMC,t}/N_{\rm BMC,r}$ at the value of
$\Gamma$ measured for the target.\\
 \noindent 3) Derive the black hole mass using the following equation:
\begin{equation}
M_{\rm BH,t}=M_{\rm BH,r}\times 
(N_{\rm BMC,t}/N_{\rm BMC,r})\times (d_t/d_r)^2 \times f_G
\end{equation}
where $M_{\rm BH,r}$ is the black hole mass of the GBH reference object
$N_{\rm BMC,t}$ and $N_{\rm BMC,r}$ are the respective BMC normalizations
for the target (\qq) and the reference source, $d_t$ and $d_r$ are the
corresponding distances, and $f_G=\cos\theta_r/\cos\theta_t$ is the
geometrical factor that depends on the respective inclination angles.

Using as a reference source GRO J1655-40, a well known microquasar whose 
parameters are tightly constrained -- $M_{\rm BH}/M_{\odot}=6.3\pm0.3$, 
$i=70^o\pm1^o$, $d=3.2\pm0.2$ kpc (\citealt{gree01,hjel95}, but see Foellmi
 2009 for a different estimate of the distance)
-- and assuming
$f_G=0.5$ which is appropriate considering that \qq\ has very likely a 
lower inclination angle compared to GRO J1655-40, we derive
$M_{\rm BH}=(0.7-2.3)\times 10^7 ~M_\odot$
($M_{\rm BH}=(2.8-6)\times 10^7 ~M_\odot$ when assuming $d_r=2$ kpc). 
The range of masses reflects the 
90\% uncertainties on the spectral parameters $\Gamma$ and $N_{\rm BMC}$.
If GX 339-4, another prototypical GBH with $M_{\rm BH}/M_{\odot}=12.3\pm1.4$, 
$d=5.8\pm0.8$ kpc \citep{shap09}, $i=45^o-70^o$ \citep{kole10}, is used 
as a reference source, assuming again $f_G=0.5$ we obtain
$M_{\rm BH}=(2-8)\times 10^7 ~M_\odot$.

The values of $M_{\rm BH}$ obtained with this new method, which is 
independent of optical measurements and of any assumption on the BLR,
appear to be in broad agreement with the values derived from the virial
theorem. Although this consistency is encouraging, the robustness of
the GBH scaling method needs to be quantitatively assessed using a
large sample of AGNs. Therefore, for the remainder of the paper we
will utilize the value determined from the H$_{\alpha}$ line and the 
virial theorem, $M_{\rm BH}\simeq7\times 10^7 ~M_\odot$ 
This corresponds to an Eddington luminosity
of $L_{\rm Edd}\simeq9\times 10^{45}$ \lum.

\subsection{Energetics of \qq}
In order to shed some light on the energetics \qq,
we will make use of some of the correlations determined by \citet{lusso10}
from a detailed optical and X-ray study of a sample of 545 X-ray selected 
type 1 AGNs from the XMM-COSMOS survey. First, by using their Equation 6, which
provides the relationship between $L_{\rm 2~keV}$ and $L_{\rm 2500\AA}$, and 
Equation 9, which describes the correlation between $\alpha_{\rm OX}$ and 
$L_{\rm 2~keV}$, we can quantitatively verify whether the UV-to-X-ray 
properties of \qq\ are consistent with the those observed in ``standard" type 
1 AGNs. By plugging the value of $L_{\rm 2500\AA}$ measured by the OM into 
Equation 6 of \citet{lusso10}, we derive $\log L_{\rm 2~keV}=25.05\pm1.16$
(where the quoted error reflects the uncertainty of the correlation parameters).
This is in full agreement with $\log L_{\rm 2~keV}=24.97$, the value 
determined from the EPIC pn measurement. Similarly, by plugging the measured value of 
$L_{\rm 2~keV}$ into Equation 9 of \citet{lusso10}, we derive 
$\alpha_{\rm OX}=1.34\pm0.65$, which again is in agreement with the
value inferred from the simultaneous measurements of the EPIC cameras and the OM
aboard \xmm, $\alpha_{\rm OX}=1.31$. 

Since in the UV-X-ray regime \qq\ behaves as a typical type 1 AGN, we can exploit 
the relationship between  the X-ray bolometric correction $\kappa_{\rm bol} \equiv 
L_{\rm 2-10~keV}/L_{\rm bol}$ and $\alpha_{\rm OX}$ presented by 
\citet{lusso10} to derive the bolometric luminosity of \qq. By plugging 
$\alpha_{\rm OX}=1.31$ into Equation 11 of \citet{lusso10}, we infer
$\kappa_{\rm bol}=17.3$, which combined with the measured value of
$L_{\rm 2-10~keV}$ leads to $L_{\rm bol}\simeq 2\times 10^{44}$ \lum. 
The bolometric luminosity in turn can be used to constrain the accretion
rate of \qq\ in terms of Eddington units. Specifically,
dividing  $L_{\rm bol}$ by the Eddington luminosity  we  determine 
$\lambda_{\rm Edd}\equiv L_{\rm bol}/L_{\rm Edd}\simeq 2\times 10^{-2}$,
which appears to be typical for ``normal" type 1 AGNs but
fairly large when compared with the values inferred for the
few confirmed true type 2 AGNs \citep[e.g.][]{bian08,panes09,shi10}. 

Finally, we can
exploit the relationship between $\lambda_{\rm Edd}$ and $\kappa_{\rm bol}$ 
from \citet{lusso10} to test our estimate of the BH mass. More specifically,
if we invert Equation (14) of \citet{lusso10} and thus express $\lambda_{\rm Edd}$
as a function of the bolometric correction $\kappa_{\rm bol}$, we derive
$\lambda_{\rm Edd}=(6\pm2)\times 10^{-2}$. This is consistent
with the value inferred above that was obtained using 
$M_{\rm BH}=7\times 10^7~M_\odot$ and hence provides an indirect confirmation
of the validity of the mass estimate.

\section{Conclusion} 
We have used data from a long \xmm\ exposure complemented with an \atca\ radio observation
and with high-quality optical spectroscopic data from the ESOFC2/NTT
to investigate in details the nature of the putative naked AGN \qq. 
The main results can be summarized as follows.

\begin{itemize}

\item  A thorough spectroscopic analysis of the NTT/EFOSC2 data reveals the presence 
of strong broad H$_\alpha$, H$_\beta$, and H$_\gamma$ lines that are in stark contrast with 
the prior classification of \citet{hawk04} based on low S/N spectra. We can
therefore rule out at high confidence level that, at the time of the NTT/EFOSC2
observation in June 2009, \qq\ was lacking a regular BLR. 

\item The lack of detection of any radio emission and the consequent low value
of the upper limit of the radio loudness implies that a relativistic jet 
(postulated to explain
the long-term optical variability observed for over two decades in \qq)
is unlikely to be present and certainly does not play a relevant role in 
the energetics of this AGN.

\item A temporal X-ray analysis indicates that \qq\ remains constant in the 
0.3--10 keV energy band on timescales of a few ks. On the other hand, a 
comparison with a \chandra\ observation carried out nearly two years before
the \xmm\ pointing reveals the presence of significant flux variability
accompanied by moderate spectral variability. \qq\ appears to follow the 
typical spectral variability trend observed in Seyfert 1 galaxies, with a 
steeper X-ray spectrum when the source is brighter.

\item A spectral analysis of the combined EPIC spectra confirms the \chandra\
results that suggested that 0.4-10 keV spectrum of \qq\ was featureless and
adequately fitted by a simple power-law. The higher quality of the \xmm\ 
spectral data makes it possible to tightly constrain the spectral parameters
and indicates that the intrinsic absorption is negligible.

\item 
From the measurement of FWHM(H$_{\alpha}$) and the virial theorem
we estimate  for \qq\ $M_{\rm BH}=(7.1\pm1.4)\times 10^7 ~M_\odot$.
Similar values are obtained by applying a new scaling technique recently 
developed for GBHs and based on the results of the X-ray spectral fit with 
a Comptonization model.

\item We use the optical-to-X-ray properties of \qq\ derived from simultaneous 
observations of the EPIC cameras and the OM aboard \xmm\ to investigate the
energetics of this source. To this end and to put \qq\ in context, we exploit the 
correlations derived by \citet{lusso10} for a large sample of AGNs
from the XMM-COSMOS survey. With a bolometric luminosity of 
$L_{\rm bol}\simeq 2\times 10^{44}$ \lum\ and an accreting rate in Eddington units
of $\lambda_{\rm Edd}\simeq 2\times 10^{-2}$ \qq\ appears to be fully consistent 
with standard type 1 AGNs and at odds with the few confirmed true type 2 AGNs.
\end{itemize}

In summary, our analysis indicates that \qq\ is not a true type 2 AGN, since the
presence of a BLR is clearly demonstrated. The relatively high accretion rate
and high X-ray luminosity of \qq\ compared to the confirmed true type 2 AGNs 
argues against an intrinsic variability scenario where the optical classification 
may change
on timescales of years. In this specific case, the original classification as
a naked AGN appears to be more likely  ascribed to a low S/N spectrum
and/or to an obscuring event during the first optical observation. 
Of the six ``naked AGN" originally selected by Hawkins (2004), three have been
observed with deep X-ray observations coupled with high S/N optical spectroscopy
and only one, Q2131-427, confirmed the genuine lack of a BLR.  
Our findings highlight
the need of high-quality X-ray and optical spectroscopic data to robustly classify
true type 2 AGNs and seem to confirm that these type of objects are indeed rare,
although a systematic analysis of a large sample of AGNs is necessary in order
to derive more general conclusions.

\begin{acknowledgments}
We thank the anonymous referee for the very constructive suggestions that 
have improved the clarity of the paper.
MG acknowledges support by the XMM-Newton Guest Investigator Program
under NASA grant NNX08AB67G and by the NASA ADP grant NNXlOAD51G. 
We thank Ranjani Sarma for helping with the X-ray spectral analysis.
\end{acknowledgments}

\end{document}